\documentstyle[preprint,tighten,aps,colordvi,epsfig,psfig]{revtex}

\begin{document}
\newcommand{\real}{{\sf I}\kern-.12em{\sf R}}
\newcommand{\fillrhd}{{\rhd\kern-.81em\bullet}\kern-.47em\triangleright}
\draft
\vskip 2cm

\title{$\Lambda$-parameter of lattice QCD with Symanzik
improved gluon actions}

\author{A. Skouroupathis and H. Panagopoulos}
\address{Department of Physics, University of Cyprus, P.O. Box 20537,
Nicosia CY-1678, Cyprus \\
{\it email: }{\tt php4as01@ucy.ac.cy, haris@ucy.ac.cy}}
\vskip 3mm

\date{\today}

\maketitle

\begin{abstract}

We compute the ratio $\Lambda_L/\Lambda_{\overline{MS}}$, where the scale
parameter $\Lambda_L$ is associated with a lattice formulation of QCD.
We consider a 3-parameter family of gluon actions, which are most
frequently used for ${\cal O}(a)$ improvement \`a la
Symanzik. The gluon action is put togeter with standard discretizations
for fermions (Wilson/clover, overlap), to provide $\Lambda_L$ for
several possible combinations of fermion and gluon actions. We employ
the background 
field technique in order to calculate the 1PI 2-point function of the
background field; this leads to the coupling
constant renormalization function, $Z_g$, at 1-loop level.

Our results are obtained for an extensive range of values for the
Symanzik coefficients.


\end{abstract}

\newpage


\section{Introduction}
\label{introduction}

The $\Lambda$ parameter of QCD has been a subject of interest for
almost three decades, since it is the necessary ``yardstick'' needed
to convert dimensionless quantities coming from numerical simulations
into measurable predictions for physical observables. 

Ever since
improved gluon and fermion actions started being employed more frequently 
in numerical 
simulations, a number of calculations of the $\Lambda$
parameter on the lattice have been carried out, using various
techniques and discretization prescriptions. Older results involving
Wilson gluons \cite{Hasenfratz}, Wilson/clover fermions
\cite{Kawai,Bode}, 
overlap fermions \cite{APV} can be found in the literature. Some
recent results regarding domain wall fermions can be found in
Ref. \cite{Aoki}. 

A calculation of the $\Lambda$ parameter which is missing is
the one involving the Symanzik improved gluon actions which are widely
used in recent simulations. The task of the present work is to fill
this gap, while at the same time we confirm some of the existing results
mentioned before. The contribution of
fermions to the computation at hand is independent
of the choice of gluon action; similarly,
gluon contributions do not depend on the fermion action.
This fact will enable us to combine our results with
previous findings regarding Wilson/clover
fermions~\cite{Bode} and overlap fermions~\cite{APV}.  

The scale parameter, $\Lambda_L$, associated with a lattice
formulation of QCD provides a relation between the lattice spacing,
$a$, and the bare coupling constant $g_{\rm o}$. It is a particular
solution of the renormalization group equation, taking the form
\begin{equation}
a\Lambda_L = \exp\left [ -\int^{g_{\rm o}}
\frac{dg}{\beta_L(g)}\right ]= \exp\left (-\frac{1}{2b_0g_{\rm
o}^2}\right ) (b_0g_{\rm o}^2)^{-b_1/2b_0} \left [ 1+{\cal O}(g_{\rm
o}^2)\right ] \label{lambdadef}
\end{equation}
\noindent where $\beta_L(g_{\rm o})$ is the lattice
$\beta$-function, and $b_0$, $b_1$ the first two coefficients of its
perturbative expansion
\begin{eqnarray}
b_0&=&\frac{1}{(4\pi)^2}\left (\frac{11}{3}N_c-\frac{2}{3}N_f\right )
\label{b0} \\
b_1&=&\frac{1}{(4\pi)^4}\left [\frac{34}{3}N_c^2-N_f
\left (\frac{13}{3}N_c-\frac{1}{N_c}\right )\right]
\label{b1}
\end{eqnarray}
($N_f$: the number of fermion flavors, $N_c$: the number
of colors.)

The $\Lambda$ parameter is a dimensionful quantity; as such it cannot
be directly obtained from the lattice. Instead, the quantity which is
calculable is the ratio between $\Lambda_L$ and the scale parameter in
some continuum renormalization scheme such as $\overline{MS}$:
$\Lambda_L/\Lambda_{\overline{MS}}$. To this end, it suffices to
compute the coupling constant renormalization function $Z_g$, relating
the bare lattice coupling $g_{\rm o}$ to the
$\overline{MS}$-renormalized coupling $g$. 

For the purposes of this calculation, we employ the background field
technique \cite{EllisBFT,WeiszBFT,Abbott}. This technique lends itself
particularly well to evaluating $Z_g$, since it obviates the need to
consider any 3-point functions.  

The rest of the paper is organized as follows: In section
\ref{Formulation} we present all the necessary background and set up. 
Our results are shown in section \ref{Results} and, finally, a
brief discussion regarding some aspects of our calculation and findings
is contained in section \ref{Discussion}.


\section{Formulation of the problem}
\label{Formulation}

We use the Symanzik improved gauge field action, involving Wilson
loops with 4 and 6 links\footnote{$1\times 1$ {\em plaquettes},
$1\times 2$ {\em rectangles}, $1\times 2$ {\em chairs} (bent rectangles),
and {\em parallelograms} wrapped around elementary
cubes.}. In standard notation, it reads~\cite{HPRSS}
\begin{eqnarray}
S_G=\frac{2}{g^2} \,\,& \Bigg[ &c_0 \sum_{\rm plaq} {\rm Re\, Tr\,}(1-U_{\rm plaq})\,
+ \, c_1 \sum_{\rm rect} {\rm Re \, Tr\,}(1- U_{\rm rect}) \nonumber \\
 & + & c_2 \,\sum_{\rm chair} {\rm Re\, Tr\,}(1- U_{\rm chair})
\,+ \, c_3 \sum_{\rm paral} {\rm Re \,Tr\,}(1-U_{\rm paral})\Bigg]
\label{gluonaction}
\end{eqnarray}

The lowest order expansion of this action (together with the gauge
fixing term, with gauge parameter $\xi$, see Eq.(\ref{gaugefix})),
leading to the gluon propagator, is \cite{Wohlert}
\begin{equation}
S_{\rm G}^{(0)} = \frac{1}{2}\int_{-\pi/a}^{\pi/a} \frac{d^4k}{(2\pi)^4}
\sum_{\mu\nu}
A_\mu^a(k)\left[G_{\mu\nu}(k)-\frac{\xi}{\xi-1}\hat{k}_\mu\hat{k}_\nu\right]
A_\nu^a(-k)\,
\end{equation}
where:\qquad \qquad $G_{\mu\nu}(k) = \hat{k}_\mu\hat{k}_\nu + \sum_\rho \left(
\hat{k}_\rho^2 \delta_{\mu\nu} - \hat{k}_\mu\hat{k}_\rho \delta_{\rho\nu}
\right)  \, d_{\mu\rho}$

\vskip 2mm
\noindent
and:\qquad\qquad $d_{\mu\nu}=\left(1-\delta_{\mu\nu}\right)
\left[C_0 -
C_1 \, a^2 \hat{k}^2 -  C_2 \, a^2( \hat{k}_\mu^2 + \hat{k}_\nu^2)
\right]$
$$ \hat{k}_\mu = \frac{2}{a}\sin\frac{ak_\mu}{2}\,, \quad
        \hat{k}^2 = \sum_\mu \hat{k}_\mu^2 \, $$
The coefficients $C_i$ are related to the Symanzik coefficients $c_i$ by
\begin{equation}
C_0 = c_0 + 8 c_1 + 16 c_2 + 8 c_3 \,, \,\,\, C_1 = c_2 + c_3\,,
\,\,\, C_2 = c_1 - c_2 - c_3
\label{ContLimit}
\end{equation}
The Symanzik coefficients must satisfy: $C_0 = 1$,
in order to reach the correct classical continuum limit.

Regarding the fermion part of the action, a variety of discretizations
are presently used in Monte Carlo simulations. The contribution of
fermions to 1 loop is independent
of the regularization chosen for the gluonic part; vice versa,
gluon contributions do not depend on the fermion action.
Consequently, the results of the present work can be directly combined with
those of previous calculations regarding Wilson/clover
fermions~\cite{Bode} and overlap fermions~\cite{APV}, yielding the
$\Lambda$ ratio for a variety of possible combinations of fermion and
gluon actions.

In the background field method, link variables are decomposed
as~\cite{EllisBFT}
\begin{equation}
U_\mu(x)=V_\mu(x)\,U_{c\mu}(x)
\end{equation}
in terms of links for a quantum field and a classical background
field, respectively
\begin{equation}
V_\mu(x)=e^{igQ_\mu(x)},\qquad
U_{c\mu}(x)=e^{iaB_\mu(x)}
\end{equation}
The $N_c \times{N_c}$ Hermitian matrices $Q_\mu$ and $B_\mu$ can be
expressed as
\begin{equation}
Q_\mu(x)=t^a\,Q_\mu^a(x),\quad B_\mu(x)=t^a\,B_\mu^a(x),\quad Tr
[t^at^b]={1\over 2}\,\delta^{ab}
\end{equation}

A choice of gauge is required for the perturbative expansion;
an appropriate gauge-fixing term is
\begin{equation}
S_{gf}={1\over {1{-}\xi}}\,\sum_{\mu,\nu}\sum_{x}Tr[D_\mu^{-}Q_\mu
  D_\nu^{-} Q_\nu]
\label{gaugefix}
\end{equation}
This term breaks gauge invariance with respect to $Q_\mu$, as it
should, but succeeds in keeping the path integral as a gauge invariant
functional of $B_\mu$. The definition of the lattice derivative, which
is covariant with respect to background gauge transformations, is
\begin{equation}
D_\mu^{-}(U_c)Q_\nu(x)=U_{c\mu}^{-1}(x-e_\mu)Q_\nu(x-e_\mu)U_{c\mu}(x-e_\mu)-Q_\nu(x)
\end{equation}

Since the quantities we will be studying are gauge independent,
we chose, for convenience, to work in the Feynman gauge, $\xi=0$.
Covariant gauge fixing produces the following action for the ghost
field $\omega$
\begin{eqnarray}
S_{gh} &=& 2 \sum_{x} \sum_{\mu} \hbox{Tr} \,
(D^+_{\mu}\omega(x))^{\dagger} \Bigl( D^+_{\mu}\omega(x) + i g_0
\left[Q_{\mu}(x), \omega(x)\right] + \case{1}{2}
i g_0 \left[Q_{\mu}(x), D^+_{\mu}\omega(x) \right] \nonumber\\
 & & \quad - \case{1}{12}
g_0^2 \left[Q_{\mu}(x), \left[ Q_{\mu}(x),
D^+_{\mu}\omega(x)\right]\right] + \cdots \Bigr),
\label{ghostaction}
\end{eqnarray}
where $D^+_{\mu}\omega(x) \equiv U_{c\mu}(x) \omega(x + {\hat \mu})
U^{-1}_{c\mu}(x) - \omega(x)$.
\vskip 1mm

Finally the change of integration variables from links to vector
fields yields a Jacobian that can be rewritten as the usual measure
term $S_m$ in the action
\begin{equation}
S_m = \sum_{x,\mu} \left \{\frac{N g_0^2}{12}\, \hbox{Tr} \,
\left (Q_\mu(x)^2 \right ) + \cdots \right \}
\label{measureaction}
\end{equation}
The measure part will not contribute to the present calculation.

In order to compute $\Lambda_L$ we need to evaluate the
renormalization function $Z_g$ for the coupling constant, up to 1
loop
\begin{equation}
g_{\rm o}=Z_g(g_{\rm o},a\bar{\mu})\,g
\label{ZgRenormFn}
\end{equation}
where $\bar{\mu}$ is the renormalization scale in the
$\overline{MS}$ scheme. Writing
\begin{equation}
Z_g(g_{\rm o},a\bar{\mu})^2=1+g_{\rm o}^2(2b_0\ln (a\bar{\mu})
+l_{\rm o}) +{\cal O}(g_{\rm o}^4) \label{Zg1}
\end{equation}
one has
\begin{equation}
l_{\rm o}=2b_0\ln\left(\Lambda_L/\Lambda_{\overline{MS}}\right)
\label{lo}
\end{equation}

To obtain $Z_g$ we only need to calculate the
one-particle irreducible (1PI) 2-point function of the background
field, $\Gamma^{(2,0,0)}(p,-p)^{ab}_{\mu\nu}$, on the lattice, to
one loop. Color symmetry and lattice rotational invariance allow one
to write~\cite{LuscherZg1}
\begin{equation}
\sum_{\mu}\Gamma^{(2,0,0)}(p,-p)^{ab}_{\mu\mu}=-3\delta^{ab}\hat{p}^2\left[1-\nu(p)\right]/g_{\rm
o}^2 \label{Gamma200}
\end{equation}
where $\nu(p)$ is a Lorentz invariant amplitude on the lattice, up
to terms which vanish as $a\to 0$; $\nu(p)$ is perturbatively
expanded as
\begin{equation}
\nu(p)=\sum_{i=1}^{\infty}g_{\rm
o}^{2i}\nu^{(i)}(p)
\label{nuDef}
\end{equation}

The background field formalism has the advantage that $Z_g$ is
directly related to the background field renormalization function
$Z_A$, through: $Z_g(g_{\rm o},a\bar{\mu})^2Z_A(g_{\rm o},a\bar{\mu})=1$. 
Consequently, no 3-point functions are needed for the evaluation of $Z_g$. 
In terms of $\nu(p)$, one can express $Z_g$ as
\begin{equation}
Z_g(g_{\rm o},a\bar{\mu})^2=1+ g_{\rm o}^{2}\left
(\nu_R^{(1)}(p/\bar{\mu})-\nu^{(1)}(ap)\right )+{\cal O}(g_{\rm
o}^4) \label{Zg2}
\end{equation}
\noindent
where
\begin{equation}
\nu_R^{(1)}(p/\bar{\mu},\xi)=\frac{N_c}{16\pi^2}\left[-\frac{11}{3}\ln\frac{p^2}{\bar{\mu}^2}+\frac{205}{36}
+\frac{3}{2(1-\xi)}+\frac{1}{4(1-\xi)^2}\right]+\frac{N_f}{16\pi^2}\left[\frac{2}{3}\ln\frac{p^2}{\bar{\mu}^2}
-\frac{10}{9}\right]
\label{nuR}
\end{equation}
is the analogous 1-loop amplitude in the $\overline{MS}$ scheme.


\section{Computation and Results}
\label{Results}

The Feynman diagrams shown in Fig. 1 contribute to $\nu^{(1)}(p)$.
All algebraic manipulation of these diagrams was performed
automatically using our software written in Mathematica. Once we
have computed $\nu^{(1)}(p)$, we use Eqs.(\ref{Zg1}) and (\ref{Zg2})
in order to obtain $l_{\rm o}$, which is the sum of a part involving
only the gluon and ghost action, and a part involving the fermion
action, i.e.,
\begin{equation}
l_{\rm o}=l_{\rm o}^g+N_f\cdot l_{\rm o}^f
\label{l0general}
\end{equation}

\begin{center}
\psfig{figure=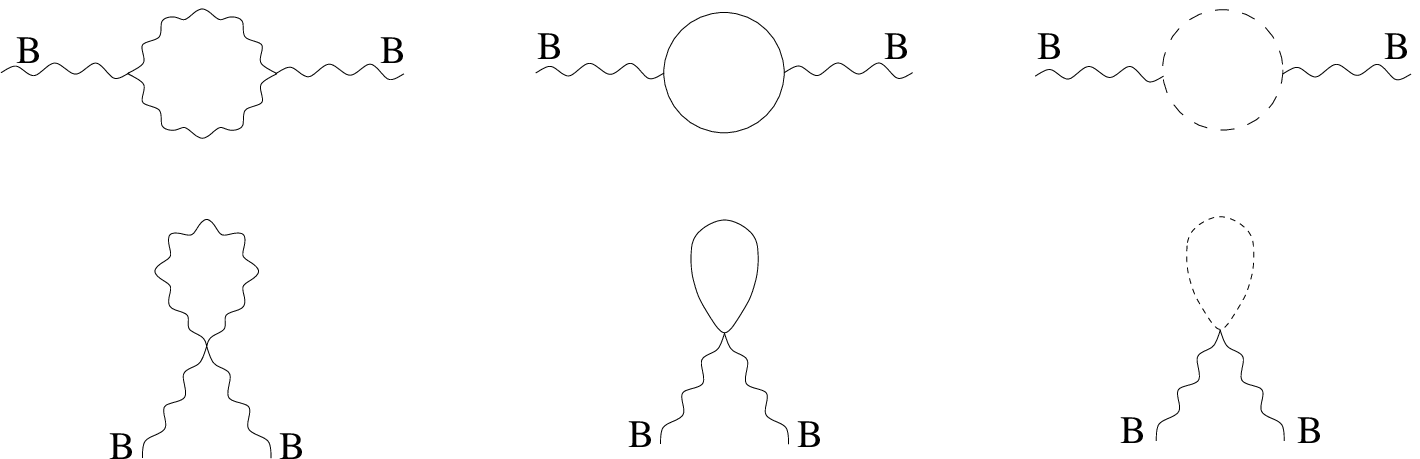,scale=1.0}
\vskip 1mm
{\small {\bf Fig. 1.}
One-loop diagrams contributing to $\Gamma^{(2,0,0)}$.
A wavy (solid, dashed) line \\
represents gluons (fermions, ghosts). The letter B stands for the
external background field.}
\end{center}

The dependence of $l_{\rm o}^g$ on the Symanzik coefficients is
rather complicated and cannot be given in closed form. However,
given that the gluon propagator depends only on the combinations
$C_1$, $C_2$ (c.f. Eqs.(\ref{gluonaction}), (\ref{ContLimit})) we
can reexpress all diagrams in terms of $C_1$, $C_2$ and one additional
parameter, say, $c_2$;
in this case the dependence on $c_2$ (at fixed $C_1$, $C_2$) is
polynomial. Thus, the part of $l_{\rm o}$ involving gluon and ghost
fields, $l_{\rm o}^g$, can be written as
\begin{equation}
l_{\rm o}^g=a_1\,\frac{1}{N_c}+a_2\,N_c+a_3\,\frac{c_2}{N_c}
+a_4\,c_2N_c+a_5\,c_2^2N_c
\label{l0gluon}
\end{equation}
where $a_i$ are numerical constants (dependent on $C_1$, $C_2$)
evaluated via numerical integration over loop momenta. We consider
ten sets of different values for the Symanzik coefficients,
corresponding to the most commonly used actions, shown in Table
\ref{paramSets}: The plaquette action, the tree-level Symanzik
improved action, the L\"uscher-Weisz tadpole improved actions (TILW),
the Iwasaki action and the DBW2 action (see
\cite{LWactions,Iwasaki,Symanzik,LWactions2,Alford,Takaishi}). The
quantities $a_i$, for each one of the ten sets of parameters, 
are presented in Table \ref{resultTab}. The variable
$c_2$ can be freely varied; $c_0$, $c_1$ and $c_3$ are then adjusted
accordingly so as to keep $C_0$, $C_1$ and $C_2$ fixed.

The fermionic part of $l_{\rm o}$, denoted by $l_{\rm o}^f$, was
calculated in \cite{APV} using 
Neuberger's overlap formulation of chiral fermions
\cite{Neuberger}, leading to
\begin{equation}
l_{\rm o}^f=-\frac{5}{72\pi^2}-k_f(\rho)
\label{l0overlap}
\end{equation}
where $k_f(\rho)$ varies from 0.07 to 0.08 in a typical range of the
overlap parameter $\rho$. For an extended list of values
of $k_f(\rho)$ see Table I of Ref. \cite{APV}.

In order to assess quantitatively the effect of the Symanzik improved 
actions on the $\Lambda$ parameter, one may consider the ratio
\begin{equation}
r_{\Lambda}\equiv
\frac{\left(\Lambda_L/\Lambda_{\overline{MS}}\right)^{Symanzik}}
{\left(\Lambda_L/\Lambda_{\overline{MS}}\right)^{Wilson}}
=\exp\left[\frac{1}{2b_0}\left(l_{\rm o}^{Symanzik}-l_{\rm
o}^{Wilson}\right)\right] 
\label{Lratio}
\end{equation}
This quantity is {\em independent} of the fermion action but still depends on
the number of flavors, $N_f$, through $b_0$. For completeness, we 
report the value of $l_{\rm o}$ found in the literature for Wilson
gluons and Wilson/clover fermions
\cite{Hasenfratz,Bode}
\begin{equation}
l_{\rm o}=1/(8N_c)-0.169955999\,N_c + N_f\,l_{01}
\label{l0Wilson}
\end{equation}
\begin{equation}
\hskip -1cm {\rm where:} \hskip 1cm
 l_{01}=0.006696001(5)-0.00504671402(1)\,c_{\rm SW}
+0.02984346720(1)\,c_{\rm SW}^2 
\label{cloverTerms}
\end{equation}
In the above, the Wilson parameter is set to $r=1$ and the clover
parameter, $c_{\rm SW}$, can be chosen arbitrarily; the dependence on
$c_{\rm SW}$ is seen to be polynomial.  

In Table \ref{LratioTab} we list the values of the ratio
$r_{\Lambda}$ for $N_f=0$ and $N_f=2$. We present $r_{\Lambda}$
for each set of parameters shown in Table \ref{paramSets}, setting
$N_c=3$ and $c_2=0$. We also list the $\Lambda$ ratio,
$\Lambda_L/\Lambda_{\overline{MS}}|_{N_f=0}$,  in the pure gauge
theory, and with two flavors of Wilson fermions,
$\Lambda_L/\Lambda_{\overline{MS}}|_{N_f=2}$. We stress that
$r_{\Lambda}(N_f=2)$ is the same for {\em all} types of fermion actions.  

In Fig. 2 we present our results for $l_{\rm o}^g$ as a function of
both $C_1$ and $C_2$, for $N_c=3$ and $c_2=0$. The range of values for
$C_1$ and $C_2$ was selected so as to encompass all values used in
current simulations.We can see that the dependence on $C_1$ is almost
linear while dependence on $C_2$ is more complicated. The
crosses correspond to the ten actions
shown in Table \ref{paramSets}. In Fig. 3 we plot the ratio
$r_{\Lambda}$ defined in Eq.(\ref{Lratio}) as a function of $C_2$.
Once again, we have set $N_c=3$, $c_2=c_3=0$ and thus $C_2=c_1$.

\vskip -1.3cm 
\hskip -5mm
\psfig{figure=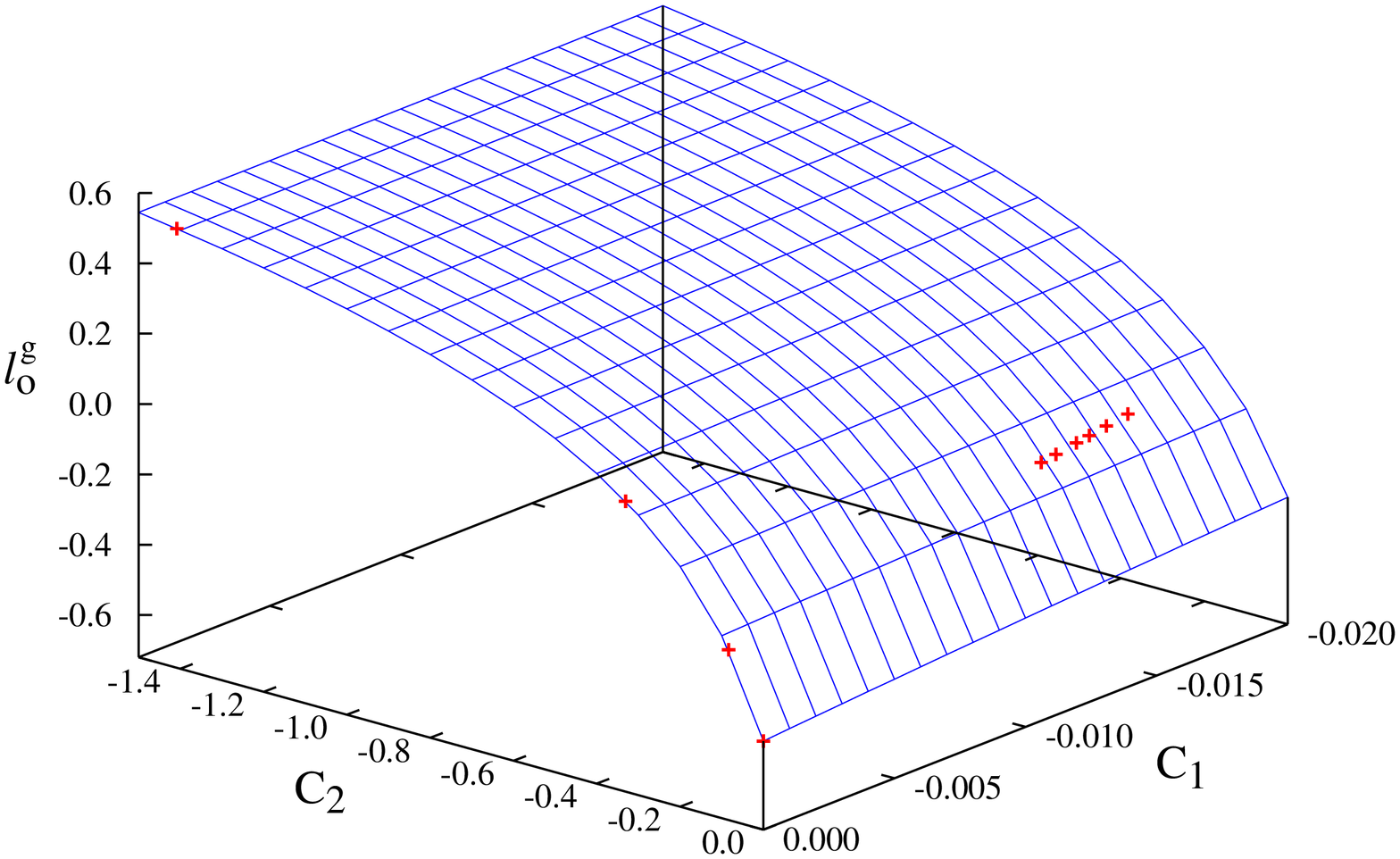,angle=-90,scale=0.63}
\vskip -5mm
{\centerline {\small {\bf Fig. 2.} $l_{\rm o}$ as a function of the parameters
$C_1$ and $C_2$ ($N_c=3$, $c_2=0$). The crosses}} 
{\centerline {\small denote the 10 set of parameters, identified by their
$C_1$, $C_2$ values, as shown in Table \ref{paramSets}.}}

\vskip 2mm
\hskip 7mm 
\psfig{figure=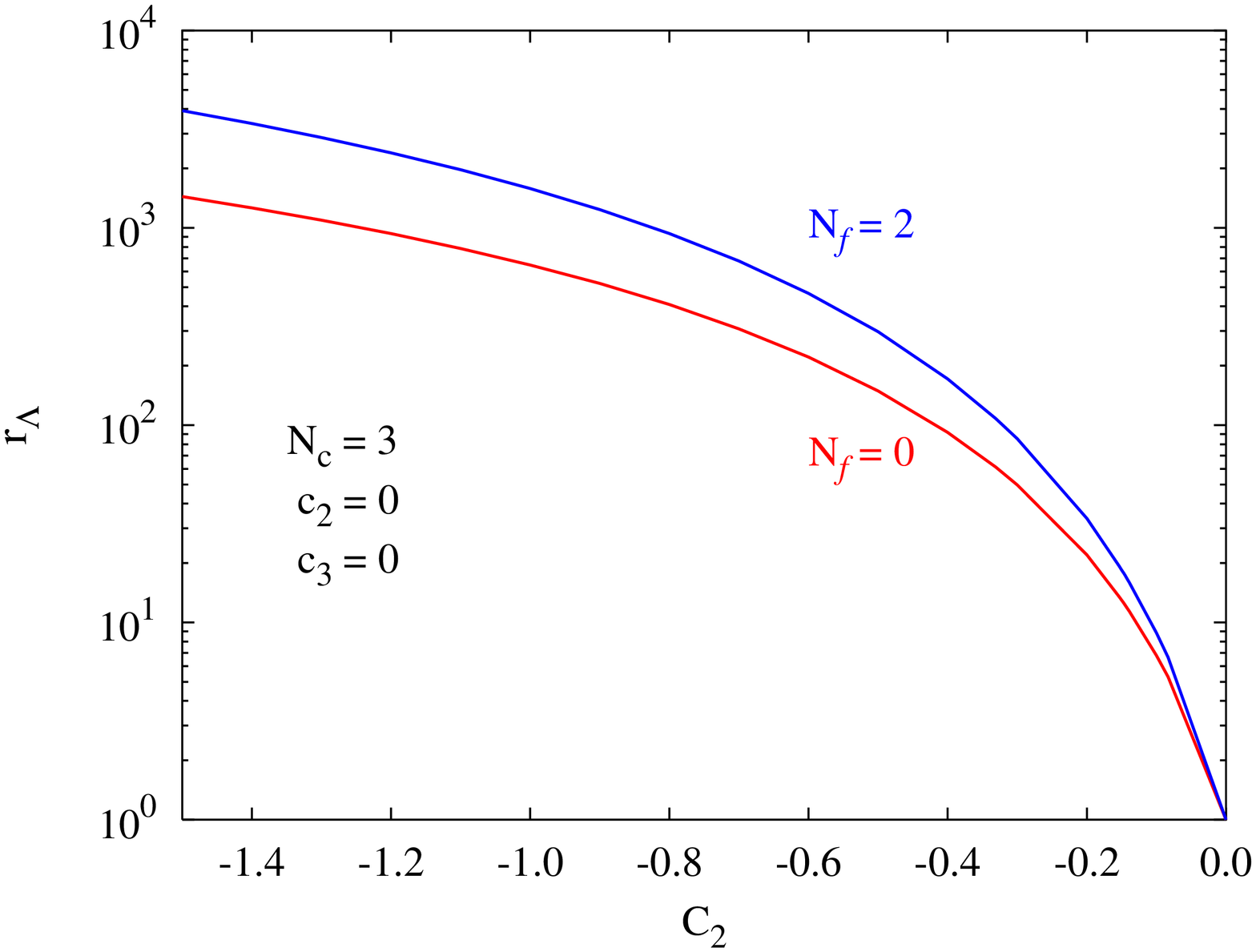,angle=-90,scale=0.5}
\vskip 2mm
\begin{center}
{\small {\bf Fig. 3.} $r_{\Lambda}$ as a function of $C_2$, for
$N_f=0,\,2$. \\
We have set $c_2=c_3=0$ (and therefore $C_2=c_1$) and $N_c=3$.}
\end{center}

For easier comparison we report some existing numbers
for the ratio
$\Lambda_L/\Lambda_{\overline{MS}}$ using Wilson
gluons and Wilson or overlap fermions (see e.g. Refs. \cite{Bode,APV}) 

$\begin{array}{lc}
{\rm Wilson\ gluons,\ Wilson\ fermions:}\hskip 2cm & \Lambda_L/\Lambda_{\overline{MS}}= 0.0243589 \\
{\rm Wilson\ gluons,\ overlap\ fermions\ (\rho=1.0):}\hskip 2cm & \Lambda_L/\Lambda_{\overline{MS}}= 0.0172702 \\
{\rm Wilson\ gluons,\ overlap\ fermions\ (\rho=1.4):}\hskip 2cm & \Lambda_L/\Lambda_{\overline{MS}}= 0.0172317 \\
\label{oldResults}
\end{array}$


\section{Discussion}
\label{Discussion}

\vskip - 2mm
In the present work we evaluated $\Lambda_L/\Lambda_{\overline{MS}}$,
for a 3-parameter family of Symanzik improved gluon actions; to this
end, we computed $Z_g$, up to 1 loop, using the background field technique.
Only diagrams with two external background fields, corresponding to
the 1PI two-point function of the background field, were involved in
the calculation, as shown in Fig. 1. Alternatively, one could study
$Z_g$ by considering the gluon-gluon-gluon, gluon-ghost-antighost or
gluon-fermion-antifermion three-point functions, together with the
self-energy diagrams for the gluon, ghost and fermion fields; of course, the
computation would be much more cumbersome in this case, due to the
complexity of the Symanzik improved actions, resulting in 
lengthier algebraic expressions. It is this very fact pointing out the
advantage of the background field technique. 

All calculations have been performed in the Feynman gauge ($\xi=0$),
and the conversion of lengthy integrands ($\sim$100,000 terms) into an
efficient Fortran 
code was carried out by our ``integrator'' program, a {\em metacode}
written in Mathematica. The numerical integrals were evaluated for
lattices up to $L=128$; the results were then extrapolated to
$L\to\infty$. Given that only a restricted set of functional forms
is sufficient to describe the behavior of the results with $L$, the
systematic error resulting 
from such an extrapolation can be estimated quite accurately. 

Special attention was given to the extraction of the dependence on the
external momentum $p$. The algebraic expressions coming from the
evaluation of Feynman diagrams were split into a logarithmically
divergent part, comprised of a limited set of tabulated lattice
integrals, and a (much larger) part which is Taylor expandable up to
second order in $p$. We have seen explicitly that terms of order
${\cal O}(p^0)$ cancel upon summation of gluon, ghost and fermion diagrams
separately, compatibly with gauge invariance.

Our results are functions of the Symanzik coefficients
$C_1$, $C_2$ and $c_2$. At fixed $C_1$ and $C_2$, the
dependence on $c_2$ is seen to be a second order polynomial, thus no
particular values of $c_2$ have to be chosen a priori; conversely, to
investigate the effect of the remaining coefficients, we selected a mesh of
25{$\times$}27 values of $C_1$, $C_2$ for numerical integration. 
The dependence on $C_1$ turns out to be almost linear, while the $C_2$
dependence is more 
complicated (see Fig. 2). 

Given that the gluon and fermion parts of the action give disjoint
contributions to $l_{\rm o}$, our present result can be directly
combined with contributions from a variety of different fermion
actions, to yield the complete effect on $\Lambda_L$.
The number of colors, $N_c$,
and the number of fermion flavors, $N_f$ can be chosen arbitrarily.    

Through Eq.(\ref{Lratio}) one can assess the effect of the Symanzik
improved actions on the $\Lambda$ parameter. All of the actions shown
in Table. \ref{paramSets}, with the exception of the DBW2 action, give
similar results (of order $10^1$) for $r_{\Lambda}$. The
most drastic effect on the $\Lambda$ parameter originates from the DBW2
improved action, where $r_{\Lambda}$ is of order $10^3$. 

\begin{table}
\begin{center}
\begin{minipage}{14cm}
\caption{The coefficients
$c_0,\,c_1,\,c_3\,(c_2=0)$, corresponding to some of the most
commonly used actions, along with the respective values for
$C_1,\,C_2$.\label{paramSets}}
\medskip
\begin{tabular}{lr@{}lr@{}lr@{}lr@{}lr@{}l}
\multicolumn{1}{c}{Action}&
\multicolumn{2}{c}{$c_0$}&
\multicolumn{2}{c}{$c_1$}&
\multicolumn{2}{c}{$c_3$}&
\multicolumn{2}{c}{$C_1$}&
\multicolumn{2}{c}{$C_2$} \\
\tableline \hline
 Set 1: Plaquette          &  1&.0        &    0&.0       &   0&.0       &   0&.0        &   0&.0      \\
 Set 2: Symanzik           &  1&.6666666  &   -0&.083333  &   0&.0       &   0&.0        &  -0&.083333 \\
 Set 3: TILW, $\beta=8.60$ &  2&.3168064  &   -0&.151791  &  -0&.0128098 &  -0&.0128098  &  -0&.138981 \\
 Set 4: TILW, $\beta=8.45$ &  2&.3460240  &   -0&.154846  &  -0&.0134070 &  -0&.0134070  &  -0&.141439 \\
 Set 5: TILW, $\beta=8.30$ &  2&.3869776  &   -0&.159128  &  -0&.0142442 &  -0&.0142442  &  -0&.144884 \\
 Set 6: TILW, $\beta=8.20$ &  2&.4127840  &   -0&.161827  &  -0&.0147710 &  -0&.0147710  &  -0&.147056 \\
 Set 7: TILW, $\beta=8.10$ &  2&.4465400  &   -0&.165353  &  -0&.0154645 &  -0&.0154645  &  -0&.149889 \\
 Set 8: TILW, $\beta=8.00$ &  2&.4891712  &   -0&.169805  &  -0&.0163414 &  -0&.0163414  &  -0&.153464 \\
 Set 9: Iwasaki            &  3&.648      &   -0&.331     &   0&.0       &   0&.0        &  -0&.331    \\
Set 10: DBW2               & 12&.2688     &   -1&.4086    &   0&.0       &   0&.0        &  -1&.4086   \\
\end{tabular}
\end{minipage}
\end{center}
\end{table}

\vskip 2.5cm

\begin{table}
\hskip -0.58cm
\begin{minipage}{17cm}
\caption{Values of the coefficients $a_1,\,a_2,\,a_3,\,a_4,\,a_5$.
Set 1 through Set 10 correspond to $C_1$, $C_2$ shown in
Table \ref{paramSets}.\label{resultTab}}
\medskip
\begin{tabular}{cr@{}lr@{}lr@{}lr@{}lr@{}l}
\multicolumn{1}{c}{Set}&
\multicolumn{2}{c}{$a_1$}&
\multicolumn{2}{c}{$a_2$}&
\multicolumn{2}{c}{$a_3$}&
\multicolumn{2}{c}{$a_4$}&
\multicolumn{2}{c}{$a_5$} \\
\tableline \hline
 1   &   0&.12499999997(6)  &   -0&.1699559990(1)  &   0&.43112525414(6)  &  -0&.0958290656(4) &  -0&.6576721162(8) \\
 2   &   0&.04217165191(7)  &   -0&.0833756545(3)  &   0&.31095652446(7)  &  -0&.031584124(1)  &  -0&.402576126(2)  \\
 3   &  -0&.0082581838(1)   &   -0&.0310360175(6)  &   0&.24637322205(3)  &   0&.000854777(6)  &  -0&.283219286(3)  \\
 4   &  -0&.010122622374(4) &   -0&.0290818603(4)  &   0&.244114928978(2) &   0&.000128847(6)  &  -0&.279256305(1)  \\
 5   &  -0&.01268965657(2)  &   -0&.0263884374(3)  &   0&.241020561259(1) &   0&.001465153(3)  &  -0&.2738512399(6) \\
 6   &  -0&.0142802781(2)   &   -0&.0247177255(2)  &   0&.23911193181(8)  &   0&.0022825409(6) &  -0&.270531792(3)  \\
 7   &  -0&.01632979450(5)  &   -0&.0225635251(5)  &   0&.2366619590(1)   &   0&.003324978(1)  &  -0&.2662885752(6) \\
 8   &  -0&.01886971906(3)  &   -0&.01989088288(9) &   0&.23364084600(6)  &   0&.004598716(5)  &  -0&.2610823949(1) \\
 9   &  -0&.07528696825(4)  &    0&.0433593330(5)  &   0&.17401920011(2)  &   0&.021865609(1)  &  -0&.159911864(1)  \\
 10  &  -0&.204424737(1)    &    0&.19876966(7)    &   0&.06102725834(6)  &   0&.03791059(2)   &  -0&.027913991(2)  \\
\end{tabular}
\end{minipage}
\end{table}

\begin{table}
\begin{center}
\begin{minipage}{16cm}
\caption{The ratio $r_{\Lambda}$ defined in Eq.(\ref{Lratio})
for each set of parameters and for $N_f=0,\,2$, along with the 
respective values for $\Lambda_L/\Lambda_{\overline{MS}}|_{N_f=0}$ 
and $\Lambda_L/\Lambda_{\overline{MS}}|_{N_f=2}$. We have set $N_c=3$ 
and $c_2=0$.\label{LratioTab}}
\medskip
\begin{tabular}{lr@{}lr@{}lr@{}lr@{}l}
\multicolumn{1}{c}{Action}&
\multicolumn{2}{c}{$r_{\Lambda}(N_f=0)$}&
\multicolumn{2}{c}{$\Lambda_L/\Lambda_{\overline{MS}}|_{N_f=0}$}&
\multicolumn{2}{c}{$r_{\Lambda}(N_f=2)$}&
\multicolumn{2}{c}{$\Lambda_L/\Lambda_{\overline{MS}}|_{N_f=2_{\phantom{2_2}}}$} \\
\tableline \hline
Set 1: Plaquette           &     1&.00000 &   0&.034711 &     1&.00000 &   0&.024359\\
Set 2: Symanzik            &     5&.29210 &   0&.18369  &     6&.65946 &   0&.16222 \\
Set 3: TILW, $\beta=8.60$  &    14&.4779  &   0&.50254  &    20&.9316  &   0&.50987 \\
Set 4: TILW, $\beta=8.45$  &    15&.0329  &   0&.52181  &    21&.8471  &   0&.53217 \\
Set 5: TILW, $\beta=8.30$  &    15&.8330  &   0&.54958  &    23&.1751  &   0&.56452 \\
Set 6: TILW, $\beta=8.20$  &    16&.3507  &   0&.56755  &    24&.0392  &   0&.58557 \\
Set 7: TILW, $\beta=8.10$  &    17&.0432  &   0&.59159  &    25&.2012  &   0&.61387 \\
Set 8: TILW, $\beta=8.00$  &    17&.9435  &   0&.62284  &    26&.7214  &   0&.65090 \\
Set 9: Iwasaki             &    61&.2064  &   2&.1245   &   107&.957   &   2&.6297  \\
Set 10: DBW2               &  1276&.44    &  44&.306    &  3423&.05    &  83&.382   \\
\end{tabular}
\end{minipage}
\end{center}
\end{table}


\end{document}